\documentclass[12pt]{article}
\usepackage{graphics}
\newcommand{\beq}{\begin{equation}}
\newcommand{\eeq}{\end{equation}}

\newcommand{\beqs}{\begin{eqnarray}}
\newcommand{\eeqs}{\end{eqnarray}}

\newcommand{\symmat}[4]{\left[ \begin{array}{cc} #1 & #2 \\ #3 & #4 \end{array} \right]}

\begin{document}
\bibliographystyle{h-physrev}

\title{Comments on Higher Loop Integrability in the $su(1|1)$ Sector of $\cal N$=4$SYM$: Lessons From the $su(2)$ Sector}
\author{A.Agarwal\thanks{abhishek@pas.rochester.edu} \\
University of Rochester. Dept of Physics and Astronomy. \\
Rochester. NY - 14627}
\maketitle

\begin{abstract}
An analysis of two loop integrability in the $su(1|1)$ sector of $\cal{N}$=4$SYM$ is presented from the point of view of Yangian symmetries. The analysis is carried out in the scaling limit of the dilatation operator which is shown to have a manifest $su(1|1)$ invariance. After embedding the scaling limit of the dilatation operator in a general (Inozemtsev like) integrable long ranged supersymmetric spin chain, the perturbative Yangian symmetry of the two loop dilatation operator is also made evident. The explicit formulae for the two loop gauge theory transfer matrix and Yangian charges are presented. Comparisons with recent results for the effective Hamiltonians for fast moving strings in the same sector are also carried out. Apart from this, a review of the corresponding results in the $su(2)$ sector obtained by Beisert, Dippel, Serban and Staudacher is also presented.
\end{abstract}
\section{Introduction and Summary:}
In the present paper, we shall carry out an analysis of higher loop integrability of the $su(1|1)$ dilatation operator of $\cal N$=4$SYM$. The analysis shall be based on the methods employed to study the integrable aspects of the $su(2)$ sector of the gauge theory dilatation operator in the recent past\cite{long-range-inz, beisert-et-al-long-range}. We shall focus mainly on the two loop integrability of the  $su(1|1)$ dilatation operator in the scaling/continuum  limit, where both $N$ and $J$ (the length of the spin chain) are allowed to be large while the ratio $\frac{\lambda }{J^2}$ is held fixed: $\lambda $ being the 't Hooft coupling of the gauge theory. The basic idea, similar to the one used in\cite{long-range-inz, Abhi-yangian}, employed in this paper would be to construct a general long ranged spin chain invariant under the Yangian of $su(1|1)$ ($Y(su(1|1))$). There is usually a lot of freedom available in the construction of long ranged spin chains with Yangian symmetries in the form of various undetermined constants. We shall fix these constants by matching them with the explicitely known  forms of the continuum limits of the   $su(1|1)$ dilatation operator. In other words we shall construct a  $Y(su(1|1))$ invariant spin chain that has the same leading order behavior (in $\frac{1}{J}$) as the gauge theory dilatation operator up to two loops. 

The interest in  the $su(1|1)$ sector is prompted in part by the fact that, on one hand it appears to be a particularly simple closed sector of the gauge theory, as is evident, for example by the fact that at one loop, the scattering matrix of the corresponding spin chain is simply the identity\cite{paba1}. Moreover, apart from the knowledge of the spin chain Hamiltonian up to three loops\cite{beisert-dyn}, the  two particle $S$ matrices and dispersion relations have been explicitely computed using the perturbative asymptotic Bethe ansatz (PABA) techniques in \cite{paba1, paba2}, to the same order in perturbation theory. Based on the assumption of factorizability of the $S$ matrix, the multi particle $S$ matrices have also been written down\cite{paba1, paba2}. This data, among other interesting things, also contains information about how the spectrum scales with the length of the spin chain $J$. Similarly, the dual string analysis, also appears to be much simpler, in this sector than in its $su(2)$ cousin; and recently, semi-classical string theory methods were employed to reproduce the famous all loop BMN\cite{bmn} (square root) formula by Stefanski and Tseytlin\cite{ts-sup-1}. However, from another point of view, this particular sector appears to be more complex than the $su(2)$ sector, and that is due to the fact that, beyond one loop, even the $su(1|1)$ symmetry of the dilatation operator is not manifest. That in turn is related to the fact that $su(1|1)$ is part of the  dynamical superalgebra, and hence, its generators also receive perturbative corrections. This technical problem, sometimes masks a lot of the simplifying features of this sector, for example, the formula for the Hamiltonian can only be written down in rather cumbersome forms, using, either a Bosonization \cite{paba1} or (through Jordan Wigner transforms) a Fermionization\cite{callan-et-al-1} of the corresponding spin chain. However, on the string theory side\cite{ts-sup-1}, the analysis is purely classical and hence free of this problem. Hence, to the extent that semi-classical string methods agree with gauge theory computations, it should be possible, at least in principle to have a manifestly $su(1|1)$ invariant language for describing the agreement between gauge theory and string theory. Moreover, the structure of Yangian symmetry is at the heart of integrability on the string theory side. Hence, it is a reasonable goal to search for $su(1|1)$ and $Y(su(1|1))$ invariance beyond the leading order on the gauge theory side of the AdS/CFT duality. In the present paper, we shall take a modest step in that direction, and see that quite like many other simplification that occur in the $su(1|1)$ sector, the next to leading order Yangian symmetry can also be realized rather explicitely. 

Integrability in $\cal N$ =4$SYM$ was first glimpsed at in the remarkable paper by Minahan and Zarembo\cite{minahan-spin}, who  showed that the matrix of anomalous dimensions/dilatation operator of $\cal N$=4$SYM$ at one loop, when restricted to the scalar sector of the gauge theory can be understood as an $so(6)$ invariant spin chain, which, amazingly enough, turned out to be exactly solvable by the methods of the algebraic Bethe ansatz. Subsequently, the full one loop dilatation operator was shown to be a  $psu(2,2|4)$ invariant supersymmetric spin chain, and its integrability was also established in\cite{beisert-et-al-super-spin, beisert-et-al-complete-one-loop}. A natural question arising from this body of work is whether or not integrability in the gauge theory persists in higher orders in perturbation theory. To this end small closed sub-sectors of the gauge theory, which are closed under dilatation have shed remarkable insights. It turns out that all  composite operators built out of two Fermions, and three complex scalars are closed under operator mixing. The dilatation operator in this closed $su(2|3)$ sector is also known in rather explicit fashion up to three loops\cite{beisert-dyn}. This sub-sector also contains two smaller closed sectors: A Bosonic one composed on two complex scalars (invariant under $su(2)\in psu(2,2|4)$) and a supersymmetric $su(1|1)$ sector composed of a complex scalar and a gaugino.  In both these small sub sectors the dilatation operator is known explicitly up to three loop level as well. Integrability in the $su(2)$ sector has been investigated vigorously in the recent past. At the one loop level, the dilatation operator in this sector takes on the form of the spin-half isotropic Heisenberg spin chain, which, of course, is integrable. As one goes higher up in the perturbation theory, the range of interactions of the spin chain increases, and it was shown in \cite{long-range-inz} that the three loop dilatation operator can be embedded in a particular long ranged spin chain (known as the Inozemtsev chain\cite{inz-rev}), after a proper redefinition of the free parameters, which are present in the Inozemtsev chain, was carried out. The Inozemtsev chain is known to commute with the generators of the $su(2)$ Yang-Baxter algebra, i.e it has $Y(su(2))$ as its hidden symmetry. The Bethe equations  (valid for chains of large length, i.e valid in the scaling limit) are also known for the Inozemtsev chain, and this translated into a knowledge of the three loop Bethe equations for the $su(2)$ sector of  $\cal N$ =4$SYM$. Matching the dilatation operator to a linear combination of the conserved charges of the Inozemtsev chain does not work beyond the three loop order, as there is a breakdown of BMN scaling in this particular long-ranged spin chain, however, the extrapolation of the three  loop Bethe equations to all loops, in a manner that is consistent with BMN scaling, was carried out in \cite{beisert-et-al-long-range}. However, whether or not the Yangian symmetries persist in this proposed all loops Dilatation operator is not clear at the moment.

In the supersymmetric $su(1|1)$ sector, the one loop dilatation operator can be diagonalized quite like the $su(2)$ case, and its underlying Yang-Baxter symmetry is  implied by that of full one loop $psu(2,2|4)$ spin chain, of which the   $su(1|1)$ spin chain is a part\cite{beisert-et-al-super-spin, beisert-et-al-complete-one-loop, paba1, paba2}. But at higher loops, although the three loop corrections to this spin chain are known, whether or not the corrections are integrable deformations of the one loop  $su(1|1)$ dilatation operator is not firmly established. However, the higher loop Bethe equations, which are expected to be exact in the scaling limit, were proposed in \cite{paba1}. One of the main objectives of this work is to establish the Yangian symmetry in the scaling limit of the two loop   $su(1|1)$ dilatation operator, which would reinforce the belief in  the exactness of the perturbative asymptotic Bethe equations in this sector\cite{paba1} at least at the two loop level. 

The relation of the exactness of the Bethe ansatz equations and an underlying Yangian symmetry can be understood as follows. There are two routes that one can possibly take towards writing down the Bethe ansatz equations for a quantum spin chain. One can either think of the Bethe ansatz, literally as an anzatz; the ansatz being that the scattering matrix be factorized. In other words, one can compute the two particle scattering matrix and make the assumption that the multi-particle scattering matrix be given as  products of the two body $S$ matrices. This approach was  used extremely successfully in \cite{paba1, paba2} in writing down the two and three loop Bethe equations, for various sector of the gauge theory, among which was also the $su(1|1)$ sector. However, as has already been pointed out in \cite{paba1}, the leap from the two body problem to the many body problem is indeed an assumption.  So the natural question is; when do the Bethe equations become exact?  This brings us to the other route to the Bethe equations which is based on $RTT$ relations and  principles of symmetry. In the many cases, where  Bethe equations have been shown to be exact solutions of  one dimensional quantum spin chains, the basic algebraic structure that has been employed is the Yang Baxter algebra, also known as the $RTT$ relations.  For example in the case of short ranged spin chains of the Heisenberg type \cite{faddeev-1, recent-open-closed-spin}, or for long ranged chains of the Haldane-Shastry type\cite{hal-shas, haldane-yan, ber-lng, ber-hal-lng} the spectrum was determined, ultimately by the $RTT$ relations, i.e the algebra obeyed by the matrix elements of the monodromy matrix. That is nothing but the Yangian algebra $Y(su(2))$(see \cite{ge-rev, bernard-yan} for reviews on this subject). Having the Hamiltonian commute with the Yangian generators was especially useful in these cases, as it enabled one to identify the Hamiltonian with one of the elements of the center of the algebra, and deduce its spectrum from a highest weight representation theory of  $Y(su(2))$\footnote{It is worth noticing that the exact routes taken towards using the Yangian symmetry to determine the spectrum is very different in the case of long ranged chains from those used in the case of short ranged ones\cite{hal-shas, haldane-yan, ber-lng, ber-hal-lng}}. In other words, there are two ways in which the Yangian algebras are  of relevance to  the study of integrable spin chains. They can either arise as symmetry algebras, as is the case in the case of the Haldane-Shastry spin chain\cite{hal-shas, haldane-yan, ber-lng, ber-hal-lng}, in which case, they give the spectrum of the Hamiltonian, along with a systematic understanding of the Higher integrals of motion\footnote{The integrals of motion for the Haldane-Shastry model were obtained from the so called quantum determinant of $Y(su(2))$\cite{int-hal-shas}.}. Though, it might well be that the $RTT$ relations that lead to the Bethe equations are generated by operators that do not commute with the Hamiltonian, i.e the Hamiltonian is a part of the algebra, but not its center. For example, in the conventional presentation of the Yangian generators, for the Heisenberg spin chain, the Hamiltonian is generated by the trace of the transfer matrix, which is not an element of its center. Though the lack of commutativity of the Yangian generators with the Hamiltonian, at least in the Heisenberg case is particularly mild, and tractable \cite{Abhi-yangian}.  However, the Hamiltonian, being a part of the algebra, once again implies that its spectrum is eventually determined by the $RTT$ relations, which are nothing but the  highest weight representation of the Yangian algebra. Although  in the present work we do not work our way towards the derivation of the spectrum of the two loop $su(1|1)$ dilatation operator, we shall nevertheless confirm its Yangian invariance in the large $J$ limit. Presumably, this algebraic structure can also be used to reproduce the PABA equations for the two loop $su(1|1)$ dilatation operator, and that remains an important possibility to explore in the direction of establishing the exactness of the higher loop PABA equations. 
The role of Yangian symmetries in some other limits of the gauge theory have also been considered in recent  papers. The role of the Yangian algebra in the weak coupling limit of the gauge theory was discussed in\cite{witten-yangian-1, witten-yangian-2}, and in the full supesymmetric planar gauge theory at one loop in\cite{dolan-yan-3}. The Yangian charges, for the $su(2)$ sector of the planar gauge theory up to two loops were found using matrix models techniques in\cite{Abhi-yangian} and  the Yangian symmetry up to three loops is in the same sector of the gauge theory is implied in\cite{long-range-inz}.  The Yangian symmetry in the scalar $so(6)$ (as well as the $su(2)$ sector) at one loop, both at finite and infinite $J$ was also shown in\cite{Abhi-yangian}.

Apart from the question about the practical utility of the Yangian symmetry in the determination of the spectrum, which clearly is a question that requires deeper study, there is also the question about whether or not this particular symmetry serves any conceptual purpose. From this point if view it is indeed an important symmetry, as this very symmetry appears in string theory on $AdS_5\times S^5$, rendering the classical world sheet sigma model integrable\cite{hidden-string-symm-1, hidden-string-symm-3, hidden-string-symm-4, yan-sigma, art-1, arsenrtt, art-2}. Although the quantum integrability of the superstring action on $AdS_5\times S^5$ is still an open issue(see \cite{frolov-et-al-baqs, berk-1, val-fl} for important developments in this direction) the classical integrability has already proven to be an extremely potent tool in carrying out  comparisons of the energies and effective Hamiltonians of fast moving Bosonic strings in various sectors of $AdS_5\times S^5$ with the gauge theory results. A matching of the string energies at one and two loops \cite{frolov-tseytlin-early, frolov-tseytlin-mul-spin, frolov-et-al-baqs, gleb-matt-1, gleb-matt-2, tseytlin-en-rec-1, tseytlin-en-rec-2} and the effective string Hamiltonians has been presented in\cite{Kruczenski-1, tseytlin-1, tseytlin-2, kristjansen-1, tseytlin-3, tseytlin-recent-review, hernandez-1, hernandez-2, ts-sup-1, ts-ncmpct}, and recently, all one loop gauge theory solutions (and two loop ones in the special $su(2)$ sector\cite{minahan-et-al-su2}) have been classified in terms of algebraic curves and matched with those arising from the string theory side\cite{beisert-alg-1, beisert-alg-2, beisert-alg-3}: the agreement  being perfect in  large $J$ limit. Integrability on the string side, is of course at the heart of a lot of the results mentioned above.  Keeping these fascinating gauge/string correspondences in mind we shall take a small step in the direction of exploring Yangian symmetries in the gauge theory, beyond the $su(2)$ sector. 

On a related note, it is worth keeping in mind that integrability in the AdS/CFT correspondence also has a M(atrix) theoretic resonance. M(atrix) theory in the plane wave background, can be regarded as a consistent truncation of the dimensional reduction of $\cal{N}$=4$SYM$ to one dimension, and it has been recently shown  to be integrable up to the fourth order in perturbation theory\cite{plefka4lp}; (see also\cite{plefka-matrix-1, plefka-matrix-3, plefka-matrix-2}).  As a matter of fact, the three loop dilatation operator of the gauge theory in the closed $su(2|3)$ matches precisely with the three loop effective Hamiltonian of the matrix model. The $su(1|1)$ sector being part of the $su(2|3)$ sector also implies, that the results presented in this paper establish  Yangian symmetry (up to two loops in the scaling limit) of the same sector of the plane wave M(atrix) theory.

The organization of the paper is as follows. In the next section, we shall recall the three loop gauge theory results in the $su(2)$ sector and review the integrable aspects of the $su(2)$ dilatation operator. Here, we shall make heavy use of the results presented in \cite{long-range-inz, beisert-et-al-long-range, Abhi-yangian}. We shall review, how the three loop dilatation operator can be embedded in the framework of $Y(su(2))$ invariant long ranged spin chains (of which the Inozemtsev chains is an example), with special emphasis on the freedom available to us in the construction of such chains in the guise of various undetermined parameters.  We shall also work out the explicit form of the monodromy matrix in this sector, up to three loops, in the large $J$ limit.  

Following this, we shall try and generalize this construction to the $su(1|1)$ sector. We shall construct a $Y(su(1|1))$ integrable spin chain which has the same leading order $\frac{1}{J}$ behavior as the two loop super Yang-Mills dilatation operator, establishing the asymptotic integrability in this sector, up to two loops. Following the comparison with the recent results of\cite{ts-sup-1}, we shall end the paper with comments about how this result might be extended to higher loops.
  
\section{The $su(2)$ Sector Revisited:}
The three loop dilatation operator of $\cal{N}$=4$SYM$ in the $su(2)$ sector is:\cite{long-range-inz,  beisert-et-al-long-range} 
\beq
D = \lambda D_1 + \lambda ^2D_2 + \lambda ^3 D_3 + O(\lambda ^4)
\eeq
where, $D_1, D_2, D_3$ are the one, two and three loop dilatation operators having the following explicit forms,
\beq
D_1 = 2\sum_i \left(I_{i,i+1} - P_{i,i+1}\right).
\eeq
\beq
D_2 = \sum_i \left(-8\left(I_{i,i+1} - P_{i,i+1}\right) + 2\left(I_{i,i+1} - P_{i,i+2}\right)\right).
\eeq
\beq
D_3 = \sum_i \left(56\left(I_{i,i+1} - P_{i,i+1}\right)  -16\left(I_{i,i+2} - P_{i,i+2}\right) -4\left(P_{i,i+3}P_{i+1,i+2} - P_{i,i+2}P_{i+1,i+3}\right)\right).
\eeq
$P_{i,j}$ is the permutation operator acting on sites $i$ and $j$.
\beq
P_{i,j} = S^a_{b}(i)S^b_{a}(j),
\eeq
where, $S^a_b(i)$ are the spin operators, $S^a_{b}(i)$ acts as $|a><b|$ at the $i$th lattice point. 
The integrability of the three loop dilatation operator can be understood as it can be expressed as a combination of the  local charges following from the most general $su(2)$ invariant transfer matrix, which can be generated by the use of Yangian symmetry. For the purposes of uncovering a similar structure in the $su(1|1)$ sector, it is worthwhile to investigate the structure of the $su(2)$ Yangian algebra and its relation to the three loop dilatation operator, in greater depth. 
\subsection{A Brief Review of $Y(su(2))$:}
The algebra, $Y(su(2))$, is nothing but the familiar algebra of the $RTT$ relations,
\beq
T_1(u)T_2(v)R_{12}(v-u) = R_{12}(v-u) T_2(v)T_1(u),
\eeq
generated by an $R$ matrix, which for $su(2)$ is nothing but,
\beq
R(\lambda ) = \lambda I - P,
\eeq
and a transfer matrix, $T(\lambda )$. In a more expanded form, the $RTT$ relations translate into the following quadratic algebra for the transfer matrix,
\beq
(\lambda - \mu ) [T^a_{b}(\lambda ), T^c_{d}(\mu)] = \left( T^a_{d}(\lambda ) T^c_{b}(\mu) -  T^a_{d}(\mu)T^c_{b}(\lambda )\right).
\eeq
Assuming the expansion of the transfer matrix in inverse powers of the spectral parameter, 
\beq
T^a_{b}(\lambda ) = \sum_n \frac{1}{\lambda ^n}(T^n)^a_{b}
\eeq
this algebra translates into the following  relations,
\beq
[(T^{n+1})^b_c, (T^{m})^a_d] - [(T^{n})^b_c, (T^{m+1})^a_d] + (T^{n})^a_c(T^{m})^b_d - (T^{m})^a_c (T^{n})^b_d =0 \label{ybe-comp},
\eeq
for the matrix elements of the transfer matrix. 
These recursion relations imply that only the first two elements $T^1$ and $T^2$ are independent and all the other $T^n, n\geq3 $ are generated by the  iterated commutators of $T^1 $ and $T^2$. Explicit formulae for all the higher $T^n$'s in terms of $T^1 $ and $T^2$ may be found in the review \cite{ge-rev}. These generators $T^1$ and $T^2$ are known as the Yangian charges, and they provide a minimal set of data required to construct the entire transfer matrix. 

While the $RTT$ relations given above, are quite universal in their form,  the Yangian generators can be many different functions of the spin operators $S^a_b$. The exact formulae for the $T^1$ and $T^2$ determine the specific details of the spin models being studied. One can ask the question about what the most general form of the Yangian generators may be, such that the transfer matrix derived from it still satisfies the Yang-Baxter algebra. This question has been addressed in the literature on  quantum spin chains\cite{haldane-yan, ber-hal-lng, ber-lng} and to anwer it, one needs to introduce the Lax operator, 
\beqs
L(i,j) &=& \theta (i,j) P_{i,j} |_{i \neq j}\nonumber \\
& &  0 |_{i = j}\label{lax}
\eeqs
This is to be regarded as the $i,j$th matrix element of the Lax operator, and no sum is implied over $i,j$. $\theta (i,j)$ is an arbitrary function of the lattice indices, which is to be determined by the condition that the transfer matrix following from this Lax operator satisfy the Yang-Baxter algebra. The transfer matrix has the following formula in terms of the Lax operator.
\beq
T(\lambda )^a_b =\sum_n \frac{1}{\lambda ^n}T^n = I +  \sum _n \frac{1}{\lambda ^{n+1}} \left( \sum_{i,j = 1}^N S^a_b(i)(L^n)_{i,j}\right)\label{transfer},
\eeq
where $L^n$ denotes the $n$th power of the Lax matrix. One can now require this transfer matrix to satisfy the Yang-Baxter algebra. The consistency condition for that is the following. One sees that there are two ways to get to $T^3$ from $T^1$ and $T^2$. There is a particular form of $T^3$, which is implied by (\ref{transfer}), and yet another one following from the Yang-Baxter algebra: setting $n=2$ and $m=1$ in(\ref{ybe-comp}), one gets,
\beq
[(T^2)^b_{a},(T^2)^d_{c}] =\delta ^d_{a}(T^3)^b_{c} -  \delta ^b_{c}(T^3)^d_{a} +\left((T^1)^b_{c}(T^2)^d_{a} - (T^2)^b_{c}(T^1)^d_{a}\right).
\eeq
These two forms of $T^3$ must be the same. This implies the following constraints on $\theta $. 
\beq
\theta (j,k)\theta (j,n) + \theta (j,k)\theta (n,k) - \theta (j,n)\theta (j,k) = \theta (j,k).
\eeq
These are known as the Serre relations, and their general solution is\cite{haldane-yan, ber-hal-lng,ber-lng},
\beq
\theta (i,j) = \left(\frac{Z_i}{Z_i - Z_j}\right),
\eeq
where, $\{Z_i\}$ are a set of arbitrary complex parameters. Thus, for any choice of the complex parameters $\{Z_i\}$, one would get a presentation of the $Y(su(2))$ algebra. Hence, it is quite evident, already at this point, that there is a great deal of freedom available to us, in the form of the undetermined constants  $\{Z_i\}$, in the construction of transfer matrices that obey the Yang-Baxter algebra.

The next question to ask is what integrable spin chain Hamiltonians such transfer matrices correspond to. To answer that we shall have to construct the center of the Yangian algebra, i.e the set of operators commuting with the Yangian charges. The charges, written in an explicit form are,
\beq
(Q^1)^a_b = (T^1)^a_b = \sum_i S^a_b(i)
\eeq
\beq
(Q^2)^a_b = (T^2)^a_b = \sum_{i,j}\theta (i,j)S^a_k(i)S^k_b(j)
\eeq
The first charge is nothing but the $su(2)$ generator, so one can look for    $su(2)$ invariant operators that commute with the second charge. As the simplest $su(2)$ invariants are given by the permutation and identity operators, one can look at operators of the form,
\beq
\sum_{ij}f_{ij}(I-P_{ij})
\eeq
and ask what the form of $f_{ij}$ has to be for such an operator to commute with $Q^2$. The requirement that such an operator commute with the Yangian generators, then fixes $f_{ij} = \theta (i,j) \theta (j,i)$, and the requirement of translation invariance forces one to set either one of the two forms for the parameters, $Z_i$\cite{haldane-yan, ber-hal-lng,ber-lng}.
\beq
Z_i = t^{-i}\label{choice},
\eeq
or,
\beq
Z_m = e^{\frac{2\pi im}{J}},
\eeq
$J$ being the length of the chain. The second form, which leads to the Haldane Shastry chain, is clearly not of interest in the present case, as it has no free parameters that one can tune to embed the dilatation operator in this chain. On the other hand,(\ref{choice}), has the undetermined parameter, $t$, that one can exploit to ones advantage. This was the reason why the Hyperbolic chain of the type (\ref{choice}) was picked in\cite{long-range-inz}, to model the three loop dilatation operator. 
Hence, the  Hamiltonian commuting with the Yangian generators is given by,
\beq
H = \sum_{i,j}-\frac{Z_iZ_j}{(Z_i - Z_j)^2}\left(P_{i,j} - I\right) = \sum_{i,j}\theta(i,j)\theta(j,i)\left(P_{i,j} - I\right).\label{hint}
\eeq
 There are other local integrals of motion for these models as well, and a general method for constructing them based on the center of the Yangian algebra, i.e the so called quantum determinant, was outlined in\cite{haldane-yan}. The first few of the local integrals are,
\beq
H_3 = \sum_{i\neq j \neq k}\theta (i,j)\theta (j,k)\theta (k,i)\left(P_{ijk} - I_{ijk}\right).
\eeq
\beq
H_4 = \sum_{i\neq j \neq k \neq l}\theta (i,j)\theta (j,k)\theta (k,l)\theta _{li}\left(P_{ijkl} - I_{ijkl}\right) -2 \sum_{i\neq j}\left(\theta (i,j)\theta (j,i)\right)^2\left(P_{ij} -I_{ij}\right).
\eeq

This is all the information necessary to embed the three loop corrected $su(2)$ dilatation operator in this general construction. To do that one would have to turn  $t$ into a  function of the 't Hooft coupling of the gauge theory i.e
\beq
t = \sum_{n=1}^{\infty } c_n \lambda ^n.
\eeq
Clearly, the case $\lim _{t\rightarrow 0}$ corresponds to,
\beq
H_{su(n)} = \sum_{i\neq j} \theta (i,j)\theta (j,i)\left(P_{i,j} - I_{ij}\right) \stackrel{\lim _{t\rightarrow 0}}{ \rightarrow } 2t\sum_{i}\left(I - P_{i,i+1}\right) + O(t^2).
\eeq
i.e we recover the one loop dilatation operator. Now allowing for $t$ to be a power series in  $\lambda $, if one keeps only the terms up to a certain order in   
$\lambda $, say $O(\lambda ^p)$  then one will have uncovered a $\lambda $ dependent perturbation of the integrable Hamiltonians given above, which have approximate Yangian invariance, i.e the Yangian generators will commute with the Hamiltonians so obtained up to terms of $O(\lambda ^{p+1})$.
\subsection{Explicit Forms of the Deformed Hamiltonians With Approximate Yangian Symmetries:}
Here we quote results accurate to the third order in $\lambda $.
\beq
t = c_1 \lambda + c_2 \lambda ^2 + c_3\lambda ^3 + O(\lambda ^4)
\eeq
\beq
(Q^1)^a_b = \sum_i S^a_b(i).
\eeq
\beqs
(Q^2)^a_b(\lambda ) &=& \sum _{i<j}\left(S^d_b(i)S^a_d(j)\right) + c_1\lambda \sum_i\left(S^d_b(i)S^a_d(i+1) - S^a_d(i)S^d_b(i+1)\right)+\nonumber \\
& & \lambda ^2\left((c_1^2+ c_2)\sum_i\left(S^d_b(i)S^a_d(i+1) - S^a_d(i)S^d_b(i+1)\right)\right)+\nonumber \\
& & c_1^2 \lambda ^2\left(\sum_i\left(S^d_b(i)S^a_d(i+2) - S^a_d(i)S^d_b(i+2)\right)\right)+\nonumber \\
& & \lambda ^3\left((c_1^3+ 2c_1c_2 + c_3)\sum_i\left(S^d_b(i)S^a_d(i+1) - S^a_d(i)S^d_b(i+1)\right)\right)+\nonumber \\
& & c_1^3 \lambda ^3\left(\sum_i\left(S^d_b(i)S^a_d(i+3) - S^a_d(i)S^d_b(i+3)\right)\right)\label{peryan}
\eeqs
\beqs
\frac{1}{2} H_{su(2), \lambda ^3} &=& \left( c_1 \lambda +( c_2 + 2c_1 ^2)\lambda ^2 +(c_3 + 4c_1c_2 + c_3 ^3)\lambda ^3\right)\sum_{i}\left( I_{i,i+1} - P_{i,i+1}\right)\nonumber \\
& &\left(c_1 ^2 \lambda ^2 + 2c_1 c_2\lambda ^3 \right) \sum_{i}\left( I_{i,i+2} - P_{i,i+2}\right)\nonumber \\
& &(c_1 ^3\lambda ^3)\sum_{i}\left( I_{i,i+3} - P_{i,i+3}\right)\label{perh}.
\eeqs
Thus we have obtained a three parameter $(c_1, c_2, c_3)$ family of deformation of the integrable $su(n)$ invariant matrix model presented previously.
The  map between perturbative analysis done so far to the radial Hamiltonian of $N$ =4 $SYM$ one needs to  identify $\lambda $ with the 't Hooft coupling of the gauge theory\cite{long-range-inz},
\beq
\lambda  = \frac{g^2_{YM}N}{16\pi ^2},
\eeq
and choose
\beq
c_1 =1, c_2 =-3, c_3 =14.
\eeq 
The expansion of the Hamiltonian, is already of the form,
\beq
H(\lambda ) = \lambda H^{(1)} + \lambda ^2 H^{(2)} + \lambda ^3 H^{(3)} +O(\lambda ^4),
\eeq
as is the expansion of the second Yangian charge. 
\beq
(Q^2)^a_b(\lambda ) = (Q^2_0)^a_b + \lambda  (Q^2_1)^a_b + \lambda ^2 (Q^2_2)^a_b,
\eeq
and the Yangian invariance up to the third order in $\lambda $ implies that
\beq
[H(\lambda ),(Q^2)^a_b(\lambda )] = O(\lambda ^4).
\eeq 
But not only, $H$, but the higher charges $H_3, H_4$ obtained from the quantum determinant or any linear combination of them will also have a similar expansion in the 't Hooft coupling, and since they also commute with the Yangian generators in this perturbative sense, one is, apart from  choosing the constants $\{c_1, c_2, c_3\}$, free to add to the Hamiltonians any linear combinations of the higher charges. This insight was used to in\cite{long-range-inz}, and  it was found that the linear combination of interest is,
\beq
I_4(\lambda ) = H_{(4)}(\lambda ) - \lambda H^{(2)},
\eeq
which for the choice of the constants $\{c_1, c_2, c_3\}$ reads as,
\beqs
I_4(\lambda ) &=& \lambda ^3 \sum_i( -8 I_{i,i+1} +8 P_{i,i+1} -4P_{i,i+2} + 4P_{i,i+3}\nonumber \\
& &+ 8\left(P_{i,i+3}P_{i+1,i+2} - P_{i,i+2}P_{i+1,i+3}\right)) + O(\lambda ^4) = \lambda ^3I_4^{(3)} + O(\lambda ^4)
\eeqs
Hence, adding everything together, we can map the perturbatively integrable model to the three loop dilatation operator $D$ as follows,
\beqs
D &=& \lambda H^{(1)} + \lambda ^2 H^{(2)} + \lambda ^3 H^{(3)}  - \nonumber \\
& &3 \lambda ^2H^{(1)} - 3\lambda ^3 H^{(2)}\nonumber \\
& &\frac{1}{2} \lambda ^3I_4^{(3)} + 22\lambda ^3H^{(1)}+O(\lambda ^4).
\eeqs

Or in more explicit terms;
\beq
D = \lambda D_1 + \lambda ^2D_2 + \lambda ^3 D_3 + O(\lambda ^4)
\eeq
where, $D_1, D_2, D_3$ are the one, two and three loop dilatation operators having the following explicit forms,
\beq
D_1 = 2\sum_i \left(I_{i,i+1} - P_{i,i+1}\right).
\eeq
\beq
D_2 = \sum_i \left(-8\left(I_{i,i+1} - P_{i,i+1}\right) + 2\left(I_{i,i+1} - P_{i,i+2}\right)\right).
\eeq
\beq
D_3 = \sum_i \left(56\left(I_{i,i+1} - P_{i,i+1}\right)  -16\left(I_{i,i+2} - P_{i,i+2}\right) -4\left(P_{i,i+3}P_{i+1,i+2} - P_{i,i+2}P_{i+1,i+3}\right)\right).
\eeq
As has been stated above, the fact that this Hamiltonian was derived from a linear combination of the charges following from the center of the Yangian algebra, automatically implies that this Hamiltonian, and consequently, the $su(2)$ sector of $\cal N$ =4 $SYM$ is integrable up to the third order in perturbation theory. 
\section{The Scaling Limit:}
The scaling limit of the dilatation operator is obtained by  taking the coherent state expectation value of the operator on a state of length $J$\cite{Kruczenski-1,  tseytlin-1, tseytlin-2, kristjansen-1, tseytlin-3, tseytlin-recent-review, hernandez-1, hernandez-2, ts-sup-1, ts-ncmpct} and letting 
\beq
\frac{1}{J}, \frac{1}{N} \rightarrow 0, \frac{\lambda}{J^2}\mbox{fixed}.
\eeq
It is not automatically guaranteed that this limit exists, for an arbitrarily chosen $Y(su(2))$ invariant spin chain. As a matter of fact, the requirement, that this scaling limit exist cuts down on the ambiguity in the choice of the parameters $c_i$ drastically. The transfer matrix is,
\beq
T^a_b(u) = I\delta ^a_b + \frac{1}{u}(Q^1)^a_b  + \frac{1}{u^2}(Q^2)^a_b +\cdots
\eeq

As mentioned previously, the other terms in the transfer matrix are generated by the iterated commutators of  Yangian charges. So, for the transfer matrix to have a well defined scaling limit, it is necessary that the Yangian charges  have a well defined continuum limit. This in turn requires one to specify the behavior of the spectral parameter $u$ as a function of $J$. We shall postulate, apart from $\frac{\lambda}{J^2}$, $\tilde u = \frac{J}{u}$ to be fixed. Specifying this behavior for the spectral parameter is necessary even for the first Yangian charge, the $su(2)$ generator, to have a sensible continuum limit,
\beq
 \frac{1}{u}(Q^1)^a_b = \sum_i\frac{1}{u}S^a_b(i) \stackrel{J\rightarrow  \infty}{=}\frac{1}{\tilde u}\int dxS^a_b(x), 
\eeq
i.e the passage from the sum over lattice indices to the integral over the continuum variable $x$ produces a factor of $J$ which has to be absorbed in a re-definition of the spectral parameter.

The one loop Hamiltonian,  Heisenberg Hamiltonian  comes with the pre factor $c_1$, which shows up in front of the two loop correction to the Yangian generator. So the two loop corrected Yangian generators are,
\beq
(Q^2)_{two-loop} = \frac{1}{\tilde u^2}\left(\int _0^1S(x)dx\int _0^xS(x')dx' - \tilde \lambda \int_0^x [S,\partial S] dx\right)
\eeq
Fixing the value of $c_1$ to be 1, and requiring that the Yangian charge have a well defined continuum limit requires that the terms of $O(\lambda ^2)$ do not have any terms involving first derivatives. This fixes 
\beq
c_2 = -3
\eeq
and the three loop corrected Yangian charges are,
\beq
(Q^2)_{three-loop} = \frac{1}{\tilde u^2}\left(\int _0^1S(x)dx\int _0^xS(x')dx' - \tilde \lambda \int_0^1 [S,\partial S] dx -  \tilde \lambda ^2 \int_0^1 [S,\partial ^3 S]dx + O(\tilde \lambda ^4) dx\right)
\eeq
These very values of $c_1, c_2$ were the ones that were mentioned in the previous section. 
Knowing these charges, allows one to find the monodromy matrix, which generates them. That is given by,
\beq
T(\tilde u, \tilde \lambda ) =  e^{\int _0^1 dx \left(\frac{1}{\tilde u}S - \frac{\tilde \lambda }{\tilde u^2}[S,\partial S] - \frac{\tilde \lambda ^2 }{\tilde u^2}[S,\partial ^3 S]\right)}
\eeq
The large $J$ limit, is to be thought of as  a classical limit, and in this limit, it suffices to replace the quantum spin matrices $S^a_b$, by classical $2\times 2$ matrices, satisfying, $S^2 = 1$. The transition from the quantum spin chain to the classical non-relativistic sigma models can best be motivated by using coherent state expansions of the spin chains, as was done in \cite{Kruczenski-1}. The dilatation operator in this classical limit assumes the form,
\beq
D = \int_0^1dx \tilde\lambda tr\left( S(x)\partial ^2 S(x) +   \tilde \lambda S(x)\partial ^4 S(x) + O(\tilde \lambda ^2)\right)
\eeq  
The classical theory, resulting from the continuum limit of the dilatation operator of course echoes the Yangian symmetry of the underlying quantum spin chain. The $RTT$ relations for the  transfer matrix of  the quantum spin chain,  translate into the following Poisson bracket relations for the monodromy matrix of the resulting classical theory,
The Poisson bracket relations satisfied by the monodromy matrix can be written down as a set of classical Yang-Baxter relations,
\beq
\{T(\lambda )\stackrel{\otimes}{,}T(\mu)\} = [r(\lambda - \mu), T(\lambda )\otimes T(\mu)],
\eeq 
where the classical 'r' matrix\cite{fadbook, bernard-yan}  is,
\beq
r(\lambda) = \frac{1}{\lambda }P.
\eeq
$P$ is the permutation operator on $V \otimes V$  \footnote{In the convention used above, the operators on the tensor product of the auxiliary vector space $V$ with itself, take on the following form in components. Let $A$ and $B$, be two $n \otimes n$ matrices, then,  
\beq
\left(A \otimes B\right)^{ik}_{jl} = A^i_jB^k_l.
\eeq
}.
The above discussion was meant to outline the basic ideas leading to classical sigma models starting from the dilatation operator of the gauge theory, and the echo of the Yangian symmetry of the quantum spin chain at the level of the sigma model. 
 
\section{The $su(1|1)$ Sector:}
Many of the integrable features of the $su(2)$ sector do generalize to the supersymmetric $su(1|1)$ sector, while others do not. The Hamiltonian in this supersymmetric sector is also know to quite high orders in perturbation theory, the three loop order to be exact, and its integrability at the one loop level is established. However, the status of higher loop integrability in this sector remains an unresolved issue. One of the  main problems in dealing with this particular sector of the gauge theory is related to the fact that  manifestly $su(1|1)$ invariant forms of the higher loop correction to the one loop dilatation operator are not known in the  $su(1|1)$ sector. That is so, because, $su(1|1)$, is part of the supersymmetry algebra, hence the $su(1|1)$ generators get $\lambda $ dependent corrections as well. Hence, the two loop Hamiltonian is $su(1|1)$ invariant only in the perturbative sense, just as the two loop $su(2)$ Hamiltonian is $Y(su(2))$ invariant, also in a perturbative sense. However, the dilatation operator has been presented in a Bosonized form  in\cite{paba1} and it was also presented in a pure Fermion language in\cite{callan-et-al-1}. In the Bosonized language, the  dilatation operator up to two loops is,
\beq
H = g^2 H_1 + \frac{g^4}{2}H_2 +O(g^6): g^2 = \frac{g^2_{YM}N}{8\pi ^2};
\eeq
where the one loop Hamiltonian is,
\beq
H_1 = (1 - \sigma _x^3) - \frac{1}{2}(\sigma _x^1\sigma _{x+1}^1 + \sigma _x^2\sigma _{x+1}^2).
\eeq
while the two loop generator is,
 \beqs
H_2 &=& \frac {1}{4}(1 - \sigma _x^3\sigma _{x+1}^3) + \frac{1}{8}(\sigma _x^1\sigma _{x+1}^1 + \sigma _x^2\sigma _{x+1}^2)-\nonumber \\
& &-\frac{1}{16}(\sigma _x^1\sigma _{x+1}^1 + \sigma _x^2\sigma _{x+1}^2)\sigma _{x+2}^3 - \frac{1}{16}\sigma _{x}^3 (\sigma _{x+1}^1\sigma _{x+1}^1 + \sigma _{x+1}^2\sigma _{x+2}^2)-\nonumber\\
& &-\frac{1}{8}\sigma _x^1(1+\sigma _{x+1}^3)\sigma _{x+2}^1 - \frac{1}{8}\sigma _x^2(1+\sigma _{x+1}^3)\sigma _{x+2}^2 \nonumber \\
& &-2\left((1 - \sigma _x^3) - \frac{1}{2}(\sigma _x^1\sigma _{x+1}^1 + \sigma _x^2\sigma _{x+1}^2)\right)
\eeqs
However, upon applying a Jordan Wigner transform, and going to the momentum space, a Fermionized version of the one and two loop Hamiltonians can be somewhat simplified,  this was the form of the Dilatation operator that was used in\cite{callan-et-al-1}.  Although, the symmetries of the Hamiltonian are no more manifest in the second language than the Bosonic, one, Fourier transforming the Hamiltonian has the advantage, that the $\frac{1}{J}$ dependence of the various terms in the Hamiltonian becomes extremely transparent. Since our  goal in this paper is only to investigate asymptotic integrability properties in the two loop Hamiltonian, the pure spinor form of the Hamiltonian is particularly useful.  The leading $\frac{1}{J}$ piece of the two loop corrected $su(1|1)$ dilatation operator as given in\cite{callan-et-al-1} is,
\beq
H^2_{Asymptotic} = 4g^2 \sum_p \sin ^2\left(\frac{p\pi}{J}\right)b^\dagger _pb_p - 8g^4 \sum_p \sin ^4\left(\frac{p\pi}{J}\right)b^\dagger _pb_p +O(g^6)\label{2lp}.\label{fer2l}
\eeq
There are other terms in the Hamiltonian as well, which are subleading order in the $\frac{1}{J}$ expansion. In other words, this two body piece of the two loop Hamiltonian is the only part that has a contribution of $O(1)$ in the asymptotic large $J$ limit. In what is to follow, we shall show that this part of the two loop Hamiltonian can be understood as a part of a hierarchy of $Y(su(1|1))$ invariant Hamiltonians. 

As mentioned in the introduction, there is a second reason to be interested in the pure spinor form of the dilatation operator. Recently, Tseytlin and Stefanski, have produced the effective semiclassical string action, also in the $su(1|1)$ sector and in the  large $J$ limit\cite{ts-sup-1}, that reproduces the famous square root BMN formula, and also matches with the leading order, $\frac{1}{J}$ results on the gauge theory side\footnote{For previous results on comparing gauge theory results to the ones obtained from the string sigma model in this sector see also\cite{hernandez-2, bellucci-fermion}}. Their formula, for the effective $su(1|1)$ string Hamiltonian is,
\beq
H_{string} = \int _0^1 \bar {\psi}\left(\sqrt{1 - \tilde \lambda ^2\partial ^2}-1\right) \psi\label{tssu11}
\eeq 
 This formula, upon discretization and  Fourier transforming to momentum space, clearly matches up with the two loop gauge theory results (\ref{fer2l}) order by order in $\tilde \lambda $, and as a matter of fact with the higher loop  (but leading order in $\frac{1}{J}$) gauge theory results, presented in\cite{paba1, callan-et-al-1} as well. Hence, working in the pure spinor picture also has the advantage of being closer  to this string theory result. This is particularly heartening, as it is well known that the full $AdS_5\times S^5$ string sigma model displays Yangian invariance, at the classical level. Hence, establishing the Yangian symmetry in the gauge theory at higher loops through a method that also makes contact with the string results allows one to get an idea of the Yangian symmetry of the string sigma model is reflected on the gauge theory side and vise versa.

\subsection{$su(1|1)$ Invariance of the Asymptotic Two Loop Hamiltonian:}

As the one loop Hamiltonian {\it is} manifestly $su(1|1)$ invariant, one can write it in terms of a graded permutation operator as, 
\beq
H^1_{su(1|1)} =  g^2\sum_x\left(I - \Pi_{x,x+1}\right),
\eeq
where, the graded permutation operator $\Pi$ has the following form in terms of the super spin operators
\beq
\Pi _{i,j} = (-1)^{\epsilon (b)}X^a_b(i)X^b_a(j).
\eeq
Super spin operators $X^a_b(i)$ satisfy the graded commutation relations,
\beq
[X^a_b(i), X^c_d(j)]_{\pm} = \delta _{ij}(\delta ^c_bX^a_d(i) - (-1)^{(\epsilon (a) +\epsilon (b))(\epsilon (c) + \epsilon (d))} \delta ^a_dX^c_b(i)).\label{su11alg}
\eeq
For the $su(1|1)$ case, $a,b = 0,1$, $\epsilon (0) = 0, \epsilon (1) = 1$.\\
To go from this particular form of the Hamiltonian, to the way it is presented as the first term in (\ref{fer2l}), one can bypass the intermediate Bosonization, and note that the $su(1|1)$ algebra (\ref{su11alg}) can be realized in a pure Fermion language using the following translation. The super spin operator, $X$, thought of as a $2\times 2 $ matrix can be written as,
\beq
X(x) = \symmat{1 - b^\dagger (x)b(x)}{b(x)}{b^\dagger (x)}{b^\dagger (x)b(x)}, [b^\dagger (x),b(x)]_+ =1.\label{purspinr}  
\eeq
Using this spinor formulation of the super spin generators, it is easy to see that,
\beq
1 - \Pi_{x,x'} = b^\dagger (x)b(x) + b^\dagger (x')b(x') - b^\dagger (x')b(x) - b^\dagger (x)b(x')
\eeq 
Hence, it is obvious that the $su(1|1)$ invariant combination of the identity and the graded permutation involves only two body terms, and in the case that $x' = x+1$, it reproduces the first term in the asymptotic Hamiltonian (\ref{fer2l}). As a matter of fact, the $su(1|1)$ invariant Hamiltonian that completely reproduces  (\ref{fer2l}), upon transformation to momentum space, is
\beq
H^2_{Asymptotic} = \sum _x \left(g^2 (1 - \Pi_{x,x+1}) + \frac{g^4}{2}\left((1 - \Pi_{x,x+2}) - 4 (1 - \Pi_{x,x+1})\right) +O(g^6)\right),
\eeq
which looks exactly like the two loop $su(2)$ Hamiltonian with the permutation operator $P$ replaced by the graded permutation operator $\Pi$. Hence, it is reasonable to expect that this Hamiltonian can be derived  from the requirement of Yangian invariance as well, which is what we shall do next.
\subsection{The Yang-Baxter Algebra For the Two Loop Hamiltonian:} The construction of $Y(su(2))$ invariant long ranged Hamiltonians can be generalized to $Y(su(1|1))$ in a rather straightforward way. We shall adhere to the approach outlined in\cite{ysu11}, where, $su(1|1)$ invariant Calogero systems were studied. The generalization to spin systems with only spin-spin interactions can be achieved by taking the limit of the system studied in\cite{ysu11}, where all the spins are frozen along their equilibrium positions. 

The spin operators (\ref{su11alg}) act on  graded vector spaces. Each vector has $m$ Bosonic and $n$ Fermionic components. Though we shall be interested only in the case $m=n=1$, it is useful to keep them arbritrary for the time being.
The $su(m|n)$ generalization of the $su(n)$ invariant Lax operator given in the previous section is given by,
\beq
L_{ij} = (1-\delta _{ij})\theta (i,j)\Pi_{ij},
\eeq
where, $\theta (i,j)$ is a function to be determined from the condition that the transfer matrix constructed from this Lax operator satisfy the graded Yang-Baxter relations; whose exact form will be given shortly. The transfer matrix is given by,
\beq
T^a_b = \sum _{ij, n}\frac{1}{u^{n+1}}X^a_b(i)L^n_{ij} = \sum_{n=0}^{\infty }\frac{1}{u^{n+1}}(T_n)^a_b.\label{sumntr}
\eeq
This transfer matrix must satisfy the graded Yang-Baxter relations. The corresponding $R$ matrix has the same form as the Bosonic one, except that the permutation is replaced by the graded permutation: 
\beq
R_{i,j}(u) = uI_{i,j} + \Pi _{i,j}.
\eeq
The formal structure of the Yang-Baxter relations remains the same,
\beq
T_1(u)T_2(v)R_{12}(v-u) = R_{12}(v-u) T_2(v)T_1(u),
\eeq
which translates into the following graded relation for the coefficients of the expansion of the transfer matrix,
\beq
[(T_n)^a_b, (T_m)^c_d]_{\pm} = \delta ^c_b (T_{m+n})^a_d - (-1)^{(\epsilon (a) +\epsilon (b))(\epsilon (c) + \epsilon (d))} \delta ^a_d (T_{m+n})^c_b
\eeq
 It may be seen from explicit computations that the general solution to the Yang-Baxter equations is provided by the same function $\theta (i,j)$ as in the Bosonic case,
\beq
\theta (i,j) = \frac{Z_i}{Z_i - Z_j},
\eeq
where $Z_i $ are arbitrary complex numbers.
Hence, the general Hamiltonian that commutes with the Yangian generators is,
\beq
H = \sum_{i,j} \theta (i,j)\theta (j,i)\left(\Pi_{i,j} - I_{ij}\right).\label{sumnh}
\eeq

Once again, one can let,
\beq
Z_i = t^{-i} 
\eeq
where $t$ can be function of the parameter $\alpha $, which will be related to the 't Hooft coupling of the gauge theory:
\beq
t = \sum_{n=1}^{\infty } c_n \alpha ^{n}.
\eeq
One can, as in the $su(2)$ case, read off the results accurate to the third order in $\lambda $.
\beqs
\frac{1}{2}H &=& \left( c_1 \alpha  +( c_2 + 2c_1 ^2)\alpha ^2 +(c_3 + 4c_1c_2 + c_3 ^3)\alpha  ^3\right)\sum_{i}\left( I_{i,i+1} - \Pi_{i,i+1}\right)\nonumber \\
& &\left(c_1 \alpha  ^2  + 2c_1 c_2\alpha  ^3 \right) \sum_{i}\left( I_{i,i+2} - \Pi_{i,i+2}\right)\nonumber \\
& &(c_1 ^3\alpha  ^3)\sum_{i}\left( I_{i,i+3} - \Pi_{i,i+3}\right).
\eeqs
The Yangian charges in this sector, obtained from the first two terms in the expansion of the transfer matrix (\ref{sumntr}), written in terms of the super spin generators  are,
\beq
(Q^1)^a_b = \sum_i X^a_b(i).
\eeq
\beqs
(Q^2)^a_b(\lambda ) &=&gr(a,b,d)(\sum _{i<j}\left(X^d_b(i)X^a_d(j)\right) + \nonumber \\
& &c_1\alpha  \sum_i\left(gr(a,b,d) X^d_b(i)X^a_d(i+1) - (-1)^{\epsilon (d)}X^a_d(i)X^d_b(i+1)\right)+\nonumber \\
& & \alpha ^2\left((c_1^2+ c_2)\sum_i\left(gr(a,b,d) X^d_b(i)X^a_d(i+1) -  (-1)^{\epsilon (d)}X^a_d(i)X^d_b(i+1)\right)\right)+\nonumber \\
& & c_1^2 \alpha ^2\left(\sum_i\left(gr(a,b,d) X^d_b(i)X^a_d(i+2) - (-1)^{\epsilon (d)}X^a_d(i)X^d_b(i+2)\right)\right)+\nonumber \\
& & \alpha ^3\left((c_1^3+ 2c_1c_2 + c_3)\sum_i\left(gr(a,b,d) X^d_b(i)X^a_d(i+1) - (-1)^{\epsilon (d)}X^a_d(i)X^d_b(i+1)\right)\right)+\nonumber \\
& & c_1^3 \alpha ^3\left(\sum_i\left(gr(a,b,d) X^d_b(i)X^a_d(i+3) - (-1)^{\epsilon (d)}X^a_d(i)X^d_b(i+3)\right)\right)
\eeqs 
where,
\beq
gr(a,b,d) = (-1)^{ \epsilon (d)(\epsilon (a)+ \epsilon (b)) + \epsilon (a)\epsilon (b)}
\eeq
After setting,
\beq
c_1 = 1, c_2 = -3,
\eeq
and forming the combination,  
\beq
I = H - 3\alpha H\label{ysu11} 
\eeq
one sees that, $I$, upon restriction to terms of $O(g^4)$, and setting $m=n=1$, is the asymptotic dilatation operator in the $su(1|1)$ sector up to two loops: $\alpha $ being related to the gauge theory coupling as $\alpha = \frac{g^2}{2}$ 
\beq
I = H^2_{Asymptotic} = \sum _x \left(g^2 (1 - \Pi_{x,x+1}) + \frac{g^4}{2}\left((1 - \Pi_{x,x+2}) - 4 (1 - \Pi_{x,x+1})\right) +O(g^6)\right).
\eeq
\subsection{Higher Loops:} If one goes to higher loops, the contribution to the asymptotic Hamiltonian is of the form,
\beq
H^3_{Asymptotic} = 32\sum_p\sin ^6\left(\frac{p\pi}{L}\right)b^\dagger_pb_b,
\eeq
which can also be expressed as a linear combination of the kind 
\beq
\left(a(1 - \Pi_{x,x+3}) + b(1 - \Pi_{x,x+2}) +c(1 - \Pi_{x,x+1})\right).
\eeq 
However, the coefficients required to do that are not consistent with what one would require to embed it into the $O(g^6)$ term in $H$: there just are not enough free parameters to be consistent with the scaling limit as well as the requirement of the precise coefficient in front of the tree loop Hamiltonian. This is the situation in the $su(2)$ sector as well(see \cite{long-range-inz} and the discussion in the previous section), which is why, one needed to add to Inozemtsev Hamiltonian, a linear combination of the higher charges to embed the three loop piece in that integrable hierarchy. Presumably, one can do that in the $su(1|1)$ case as well, and we hope to report some results in that direction in the near future. However, it is worth bearing in mind that the violation of the BMN scaling at the four loop level, was why one needed to invent a new long ranged spin chain\cite{beisert-et-al-long-range} to describe the $su(2)$ Hamiltonian beyond three loops, and such problems might occur in the $su(1|1)$ sector as well. However, as we are working directly in the scaling limit; and it is known that in this limit classical string theory works to all orders\cite{ts-sup-1}, it might be worth investigating if  all loops Yangian symmetry in scaling limit of the $su(1|1)$ can be understood in a transparent way from that of the world sheet sigma model. 

\subsection{The Scaling Limit:}
In this final section, we shall comment on the derivation of the (classical) continuum limit of the spin chains via super spin coherent states, as was done recently in \cite{ts-sup-1}. 

As in the Bosonic case, the coherent states are of the form,
\beq
|Z> = Z^{i_1}(1)Z^{i_2}(2)\cdots Z^{i_{J}}(J)|i_1i_2\cdots i_{J}>,
\eeq
except that the functions $Z^{k}(x)$ are anticommuting for $k=1$ and commuting for $k=0$. More precisely\cite{ts-sup-1},
\beq
Z^i = (\phi,\psi), Z_i = (\phi ^\star , \bar{\psi}).\label{cpn}
\eeq
$\phi $ is the (complex) scalar degree of freedom while, $\psi $ is the Fermionic one.
Normalization of the coherent states enforce the constraints, 
\beq
Z_iZ^i = Z^iZ_i = 1, 
\eeq
where the convention, as employed in\cite{ts-sup-1}, is,
\beq
Z^iZ_i = Z^0Z_0 - Z^1Z_1, Z_iZ^i = Z_0Z^0 + Z_1Z^1.\label{cons}
\eeq   
The basic idea is now to take the coherent state expectation value of the Dilatation operator in these states and treat the expectation value, as a classical Hamiltonian. 
\beq
H^1_{classical} = <Z|H^1_{su(1|1)}|Z>
\eeq
Explicit computations give,
\beq
H^1_{classical} = - g^2\sum_x\left(Z^i(x)Z_j(x)Z^j(x+1)Z_i(x+1)\right)
\eeq
One can introduce the (super)matrix,
\beq
M^a_b = Z^aZ_b,
\eeq
in whose terms the classical Hamiltonian assumes a manifestly $su(1|1)$ invariant form,
\beq
H^1_{classical} = - g^2\sum_x\left(str(M(x)M(x+1)\right),
\eeq
which in the large $J$ limit becomes,
\beq
H^1_{classical} = -\frac{1}{2}\tilde g^2 \int_0^1\left(str(M(x)\partial ^2_xM(x)\right),
\eeq
where, $\tilde g = \frac{g}{J}$
The normalization constraints (\ref{cons}), show up as constraints on the matrix $X$ as,
\beq
M^2 = M, strM = 1.
\eeq 
These should in general be imposed using Lagrange multipliers, or alternatively, one can solve for these constraints and rewrite the Hamiltonian in terms of the degrees of freedom left over after the constraints are solved. This is equivalent to gauge fixing the classical Hamiltonian. Indeed it can be written down, when expressed in terms of the $Z's$ as a supersymmetric $CP^{m|n}$ model\cite{ts-sup-1}, which has a gauge symmetry. The upshot of this symmetry is that the Bosonic degree of freedom can be gauged away, allowing one to make contact with the pure Fermionic form of the  $su(1|1)$ Hamiltonian. More precisely, the constraints can be solved as,
\beq
\phi = \left(1 - \frac{1}{2}\bar{\psi}\psi\right).
\eeq
Using this solution, the matrix $X$ becomes,
\beq
M = \symmat{1 - \bar{\psi}\psi}{ {\psi}}{\bar \psi}{- \bar{\psi}\psi},
\eeq 
which is the coherent states method of reproducing (\ref{purspinr}). $\psi, \bar \psi$ are grassman numbers, they are the classical analogues of $b, b^\dagger $. The classical Hamiltonian now takes on the following form,
\beq
H^1_{classical} = -\frac{1}{2}\tilde g^2 \int_0^1dx \left(str(M(x)\partial ^2_xM(x)\right) = -\tilde g^2\int_0^1dx\left(\bar{\psi}\partial^2 \psi \right),
\eeq
Similarly, 
\beq
H^2_{classical} = -\frac{\tilde g ^4}{2}\int_0^1dx \left(\bar{\psi}\partial^4 \psi \right).
\eeq
These are the first two terms in the expansion of (\ref{tssu11}), showing the asymptotic, two loop agreement of the gauge theory results with the effective string sigma model. 

As in the $su(2)$ case, the scaling limit is a classical limit, with the quantum super spin operators, being replaced by classical $2\times 2 $ super matrices\footnote{Keeping with the conventions of\cite{ts-sup-1} the upper and lower indices in the classical limit are the reverse of those in employed in the quantum spin chain. The reversal takes place as the coherent state indices are regarded as the 'upper' ones in (\ref{cpn}). i.e $(X^a_b)_{spin-chain}\rightarrow (M^b_a)_{sigma-model}$}. Hence, one can now formulate the integrability of this resulting two dimensional field theory also in terms of (semi-classical) Yangian symmetries.  As was outlined in the discussion on the scaling limit of the two loop $su(2)$ dilatation operator, one can now relate the scaling limit of the Yangian charges and obtain the classical Yangian charges, that commute with the two loop scaling Hamiltonian,
\beq
H_{classical} = \int_0^1dx\left(\tilde g^2 str(M(x)\partial ^2M(x)) + \frac{\tilde g^4}{2} (M(x)\partial ^4M(x)) + O(g^6)\right) 
\eeq
After rescaling the spectral parameter $u \rightarrow \tilde u = \frac{u}{J}$, the continuum limits of the Yangian charges may be taken as well. They are,
\beq
(Q^1)^a_b = \int _0^1dx M^a_b(x)
\eeq
\beq
(Q^2)^a_b = \int _0^1dx M^d_b(x)\int _0^xdx' M^a_d(x') + \frac{1}{2}\tilde g^2  \int _0^1dx \left(\partial M^d_b(x) M^a_d(x)- M^d_b(x)\partial M^a_d(x)\right),
\eeq
with the convention, that the sum from the upper left hand corner to the lower right hand corner is signed, i.e $M^d_bM^a_d $ is a short hand notation for $(-1)^{\epsilon (d)}M^d_bM^a_d$. The monodromy matrix, (to $O(g^2)$) can be written down as,
\beq
T(\tilde u) = e^{\frac{1}{\tilde u}\int_0^1dx\left(M + \frac{1}{2\tilde u}\tilde g^2 ((\partial M)M - M(\partial M)) + O(\tilde g ^4)\right)}.
\eeq
In the case of the $su(2)$ sector, the modnodromy matrix, and indeed the Yangian charges, for the semiclassical coherent state sigma models can be obtained as particular limits of those corresponding to the $S^3$ sectors of classical strings on $AdS_5\times S^5$. In the $su(1|1)$case, the effective semi-classical string action consistent with the all loop $BMN$ behavior of the asymptotic  $su(1|1)$ dilatation operator was derived from the full classical superstring action in \cite{tseytlin-3}. Hence, in principle, the non-local conserved (super) charges, given above, and indeed the monodromy matrix should follow from a similar expansion of the charges and classical monodromy matrix of the dual classical string theory. That analysis should also inform us on how the Yangian symmetry is realized to all loops (in the asymptotic large $J$ limit) in the gauge theory, paving the way for an all loop extension of the above results. We hope to report on this interesting possibility in the near future.  
    
\section{Appendix:}
In this appendix we shall demonstrate the commutation of the two loop Yangian generator with the two loop Hamiltonian. 

The tow loop corrected $su(1|1)$ Hamiltonian is,
\beq
H = g^2 H^{(1)} + g^4 H^{(2)} +O(\lambda ^3) .
\eeq
Where,
\beq
H^{(1)} =  \sum_x\left(1 - \Pi_{x,x+1}\right),
\eeq
and,
\beq
H^{(2)} =  \frac{1}{2}\sum_x\left((1 - \Pi_{x,x+2}) - 4(1 - \Pi_{x,x+1})\right).
\eeq
Similarly, the two loop Yangian charges are,
\beq
Q^a_b(\lambda) = (Q^{(1)})^a_b + \frac{g^2}{2} (Q^{(2)})^a_b + O(g^4),
\eeq
where,
\beq
(Q^{(1)})^a_b = (-1)^{ \epsilon (d)(\epsilon (a)+ \epsilon (b)) + \epsilon (a)\epsilon (b)}( \sum _{i<j}\left(X^d_b(i)X^a_d(j)\right),
\eeq
and
\beq
(Q^{(2)})^a_b = \sum_i\left((-1)^{ \epsilon (d)(\epsilon (a)+ \epsilon (b)) + \epsilon (a)\epsilon (b)}X^d_b(i)X^a_d(i+1) - (-1)^{ \epsilon (d)}X^a_d(i)X^d_b(i+1)\right).
\eeq
To establish perturbative integrability of the two loop Hamiltonian, the necessary condition is,
\beq
[H^{(1)}, (Q^{(2)})^a_b] + [H^{(2)},(Q^{(1)})^a_b] = [\sum_x \Pi_{x,x+1}, (Q^{(2)})^a_b] + [\sum_x \Pi_{x,x+2}, (Q^{(1)})^a_b] = 0\label{haus}
\eeq
Explicit computations give,
\beqs
\sum_x \Pi_{x,x+2}, (Q^{(2)})^a_b] &=& (-1)^{ (\epsilon (l)+\epsilon (n))(\epsilon (b)+\epsilon (n))}\Upsilon ^{aln}_{nbl}\nonumber\\
& & + (-1)^{ (\epsilon (a)\epsilon (b) +\epsilon (n)\epsilon (b) + \epsilon (a)\epsilon (n) + \epsilon (l) )}\Upsilon ^{nal}_{bln}\nonumber\\
& & - (-1)^{ (\epsilon (n)+ \epsilon (a))(\epsilon (n) + \epsilon (l))}\Upsilon ^{lan}_{nlb}\nonumber \\
& & - (-1)^{ \epsilon (n)(\epsilon (a) + \epsilon (b)) + \epsilon (a)\epsilon (b) + \epsilon (l) }\Upsilon^{nla}_{lbn}\nonumber.
\eeqs
We have used the short hand notation for translationally invariant combinations of spin operators,
\beq
\Upsilon ^{i_1i_2\cdots i_n}_{j_1j_2\cdots j_n} = \sum_iX^{i_1}_{j_1}(i)X^{i_2}_{j_2}(i+1)\cdots X^{i_n}_{j_n}(i+(n-1)).
\eeq
Similarly, one also has,
\beqs
\sum_{x,i}[\Pi_{x,x+1},(-1)^{ \epsilon (d)(\epsilon (a)+ \epsilon (b)) + \epsilon (a)\epsilon (b)}\left(X^d_b(i)X^a_d(i+1)\right)]&=& (-1)^{ \epsilon (n)(\epsilon (a) + \epsilon (b)) + \epsilon (a)\epsilon (b) + \epsilon (l) }\Upsilon^{nla}_{lbn}\nonumber \\
& &- (-1)^{ (\epsilon (a)\epsilon (b) +\epsilon (n)\epsilon (b) + \epsilon (a)\epsilon (n) + \epsilon (l) )}\Upsilon ^{nal}_{bln} \nonumber
\eeqs
\beqs
-\sum_{x,i}[\Pi_{x,x+1},(-1)^{ \epsilon (d)}\left(X^a_d(i)X^d_b(i+1)\right)] &=& -(-1)^{ (\epsilon (l)+\epsilon (n))(\epsilon (b)+\epsilon (n))}\Upsilon^{aln}_{nbl}\nonumber\\
& &(-1)^{ (\epsilon (n)+ \epsilon (a))(\epsilon (n) + \epsilon (l))}\Upsilon ^{lan}_{nlb}\nonumber 
\eeqs
These results establish that (\ref{haus}) is indeed satisfied.

{\bf Acknowledgments:} It is a pleasure to thank Sergey Frolov, Thomas Klose, Arsen Melikyan, Jan Plefka, Sarada Rajeev and Radu Roiban for stimulating discussions and Gleb Arutyunov, Niklas Beisert, Thomas Klose, Jan Plefka and Radu Roiban for taking the time to read through an earlier version of this manuscript. The wonderful hospitality of Albert Einstein Institute (Golm), where this paper was written, is gratefully acknowledged. This research was supported in part by US Department of Energy grant number DE-FG02-91ER40685.

\bibliography{abhishekbib}
\end{document}